\documentclass[12pt]{article}
\usepackage{graphicx}
\usepackage{epsfig}
\usepackage{overcite}
\usepackage{color}
\usepackage{longtable}
\begin{document}

\begin{center}

\noindent {\LARGE {\bf Symmetry, Levitation Effect and Size Dependent Diffusivity Maximum
}}

\noindent{\bf Manju Sharma$^1$, S. Yashonath$^{1}$}\\                                                                                                                                                                                                    \baselineskip=12pt                                                                 \noindent {\it $^{1}$ Solid Sate and Structural Chemistry Unit}\\                  \noindent {\it Indian Institute of Science, Bangalore, India - 560 012}\\          \end{center}                                                                                                                                                          \baselineskip=12pt                                                                                                                                                    \vspace*{1.0cm}                                                                    \begin{center}                                                                     \large {\bf Abstract}                                                              \end{center}                                                                                                                                                                        

Diffusion invariably involves motion within a medium.
An universal behavior observed is that self diffusivity exhibits a maximum
as a function of the size of the diffusant when the diffusant is confined to a medium,
as a result of what is known as the Levitation Effect. Such a maximum in self diffusivity
has been seen in widely differing medium : microporous solids, dense liquids and close-packed
solids, ions in polar solvents, etc. The effect arises because the forces exerted on the
diffusant by the medium in which it is confined is a minimum for the size of the diffusant
for which self diffusivity is a maximum. We report here simulations on a diatomic species
confined to the cages of zeolite Y. Several different simulations in which the two
atoms of the model diatomic species interact with equal strength(example, $O_2$, the symmetric case) and
with unequal interaction strengths (example, $CO$, asymmetric case) are modeled here. Further, the
bond length of the diatomic species is varied. Our results for the symmetric case shows that
self diffusivity is maximum for a large enough bond length which fits snugly into the 12-ring window
of zeolite Y. For weakly asymmetric case, a weak maximum is seen as a function of the
bond length of the diatomic species. However, for strongly asymmetric case, no maximum in self
diffusivity is seen for all the bond lengths studied. This demonstrates
close relation between symmetry and the diffusivity maximum and provides a direct evidence for the
need of force cancellation to observe the Levitation Effect.

\baselineskip=20pt

\section {Introduction}

\begin{figure}
\begin{center}
{\includegraphics*[width=10cm]{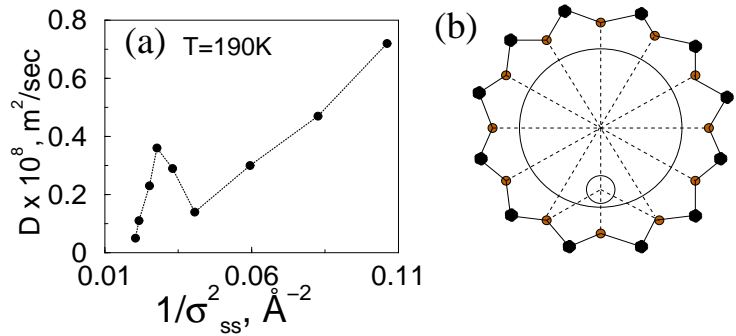}}
\caption{(a) Variation of self diffusivity as a function of the  
diameter of a spherical sorbate in zeolite-Y at 190K \cite{yashosanti94b}. (b) Schematic 
representation of force 
acting when sorbate is small relative to the bottleneck and when it is comparable to the
bottleneck. In the latter case, the force from diagonally opposite directions are equal
and opposite leading to negligible net force on the sorbate.}
\label{le}
\end{center}
\end{figure}

Size dependence of self diffusion has attracted considerable 
attention over more than a hundred years.\cite{Born, Bagchi_ACR} Experimental ionic 
conductivities in solution with water, acetonitrile and other
solvents show an anomalous dependence on ionic radii with Cs$^+$ 
having higher conductivity than Li$^+$ in most solvents. 
Early theories by Max Born \cite{Born} were followed by continuum-based theories
by Hubbard and Onsager \cite{HO_theory} and Zwanzig \cite{Zwanzig_1963, Zwanzig_1970} and 
more recently, microscopic 
theories \cite{Wolynes_1980, Bagchi_ACR, Chen_Adelman, Chandra_Chowdhury, Rasaiah}. 
Very recently, work from this laboratory suggested Levitation 
Effect as a possible reason for the higher conductivity of the larger ions \cite{pradip_lynden_ion}. 
The Levitation Effect refers to the anomalous maximum in self diffusivity 
on the size of the diffusant when the diffusant has a diameter that is 
comparable to the narrowest part of the void space through which it has to 
pass through during diffusion. 
Typical variation of self diffusivity with 
the size of the diffusant is shown in Figure \ref{le}(a). The effect of forces due to the
host medium on the different sizes guests diffusing through the narrowest region host is
shown in Figure \ref{le}(b).  
The diffusivity maximum
is a generic feature that is ubiquitous and has been seen in a wide variety 
of phases : porous crystalline solids, dense amorphous solids, dense liquids, 
ions in water and other solvents. 

In spite of many investigations into Levitation Effect, more detailed 
understanding is still lacking into the origin of this effect.

The diffusivity of diffusant decreases with increase in 
size as predicted by kinetic theory and Stokes-Einstein relationship. In Figure \ref{le}
very small sorbate sizes show such decrease in diffusivity with size and
diffusivity is inversely proportional to the square of the size. This
regime is termed as linear regime(LR). But when the size of diffusing species is 
similar to the narrowest region of the void(window in microporous solids, neck in
phases with nanosized voids in condensed media) present in the host system through
which it is diffusing, it shows enhanced diffusivity or Levitation effect. 
In Figure \ref{le} the sorbates which show increase in diffusivity with increase
in size belong to anomalous regime(AR) and in this regime the sorbate with size
similar to the window or neck region shows diffusivity maximum. 
To quantify the
Levitation effect observed in different systems, a parameter called {\it levitation
parameter}, $\gamma$ is defined and it is the ratio of the optimum size of the
diffusant and the narrowest region of the void in the medium.

\begin{equation}
\gamma=\frac{2\times 2^{\frac{1}{6}}\sigma_{opt}}{\sigma_w}
\label{lev_par}
\end{equation}

Optimum size is when the size of diffusant is similar to the window region and thus, 
diffusant-host interactions are optimum, $\gamma$ $\equiv$ 1 and the diffusant
hows anomalous diffusion or Levitation Effect. In case of disordered solids
$\gamma$ $<$ 1. When the size of diffusant is very small than the window of zeolite,
it feels net attraction towards one part of the window as can be seen in Figure \ref{le}(b)
and thus perform slow diffusion. When an optimum size diffusant passes through the
window of the host medium, the net forces acting on it due to the host atoms cancel out
due to the similar size of diffusant and window and it acts like a weakly bound particles
with a high diffusivity. 

In this work, we report a molecular dynamics simulation of a diatomic model 
guest AB sorbed within zeolite NaY. We investigate the effect of bond length, 
$d$, of A-B on self diffusivity $D$ of the diatomic molecule. All the results are 
reported in terms of a parameter, $l$ which is half of the guest bond length, $d$.
We find the expected anomalous maximum $D(l)$ of self diffusivity on $l$ when 
the bond length A-B, $d$ is comparable to the window diameter of the 12-membered 
ring, the bottleneck for diffusion. Further, we report the effect of symmetry 
in the behavior of the curve $D(l)$. The result demonstrates the close 
relationship between symmetry and the Levitation Effect.

\section{Methods}

\subsection {Model and intermolecular potential}
 
The simulation cell consists 2$\times$2$\times$2 unit cells of zeolite NaY.  
The initial configuration has three diatomic (AB) guest molecules per cage. 
The unit cell coordinates of zeolite Y are taken from Fitch and 
coworkers \cite{Fitch}. Interaction between the diatomic species and the 
zeolite are accounted for by site-site interaction between the two sites 
A, B and zeolite atoms Si, O and Na. The interactions are through 
Lennard-Jones potential without any long-range forces. The united atom parameters of 
ethane given by Jorgensen have been employed in the present work with suitable 
modification \cite{Jorgensen}. The parameters for guest-zeolite interaction
have been obtained from the Lorentz-Berthelot combination rules. The zeolite 
parameters have been taken from the previous work of Kiselev and 
coworkers \cite{Kiselev}. The zeolites-zeolite parameters are :
($\sigma_{OO}$= 2.5447\AA, $\epsilon_{OO}$= 1.2891kJ/mol, 
$\sigma_{NaNa}$= 3.3694\AA\ and $\epsilon_{NaNa}$=0.0392kJ/mol).

We have carried out simulations with (a) $\epsilon_A$ = $\epsilon_B$ 
(symmetric case or sym) (b) $\epsilon_A$ $\neq$ $\epsilon_B$  (intermediate 
asymmetry or i-asym) (c) $\epsilon_A$ $>>$ $\epsilon_B$ (extreme asymmetry 
or e-asym). The corresponding potential parameters for A site of diatomic 
species with O and Na of zeolite were derived from the combination rule. 
The precise values of the interaction parameters employed in the present 
study for symmetric case are listed in Table \ref{potpar_sym}. For the 
intermediate as well as extreme asymmetric cases, we derived the 
$\epsilon_{Ah}$ parameters as follows :

\noindent
$\epsilon^{i-asym}_{xNa}$ = $\epsilon^{sym}_{xNa}$ $\pm$ 0.15   

\noindent
$\epsilon^{i-asym}_{xO}$ = $\epsilon^{sym}_{xO}$ $\pm$ 0.75

\noindent
$\epsilon^{e-asym}_{xNa}$ = $\epsilon^{sym}_{xNa}$ $\pm$ 0.25   

\noindent
$\epsilon^{e-asym}_{xO}$ = $\epsilon^{sym}_{xO}$ $\pm$ 1.5 \hspace{3cm} $x = A,B$

The total interaction energy is 

\begin{equation}
U_{tot} = U_{gh} + U_{gg}
\end{equation}
\label{inter_eqn_symm}

\begin{table}
\caption {Lennard-Jones interaction parameters for the symmetric guest and host.}
\begin{center}
\begin{tabular}{|c|c|c|c|c|}\hline
{type of interaction}&$\sigma$, \AA\ &$\epsilon^{sym}$, kJ/mol & $\epsilon^{i-asym}$, kJ/mol & $\epsilon^{e-asym}$, kJ/mol \\\hline
AA & 3.78 & 0.867&0.867&0.867\\
AO & 3.16235 & 1.5858&0.835805 &0.0858050\\
ANa & 3.57465 & 0.2766&0.126636 &0.0266360\\
BB & 3.78 & 0.867 &0.867 &0.867 \\
BO & 3.16235 & 1.5858 &2.33580 &3.08580\\
BNa & 3.57465 & 0.2766 &0.426636 &0.526636\\\hline
\end{tabular}
\label{potpar_sym}
\end{center}
\end{table}

\subsection {Computational Details}

All the simulation runs are made in microcanonical ensemble with DLPOLY 
package using Verlet leapfrog integration scheme \cite{dlpoly}.
The zeolite atoms are kept frozen during the simulation. A timestep of 1fs 
gives a good total energy conservation of the order of 10$^{-5}$. A cut off 
radius of 20 \AA\ is used to calculate guest-guest and guest-host 
interactions. 
All the runs have been made at a temperature of 200K and the bond length 
of ethane molecule is varied in the range of 1.54 to 4.0\AA\ in increments 
of 0.2\AA. An equilibration run of 500ps with a production run of 1ns 
has been made. The position coordinates, velocities and forces of guests are 
stored at an interval of 25fs.

\section{Results and Discussion}

\begin{figure}
\begin{center}
{\includegraphics*[width=10cm]{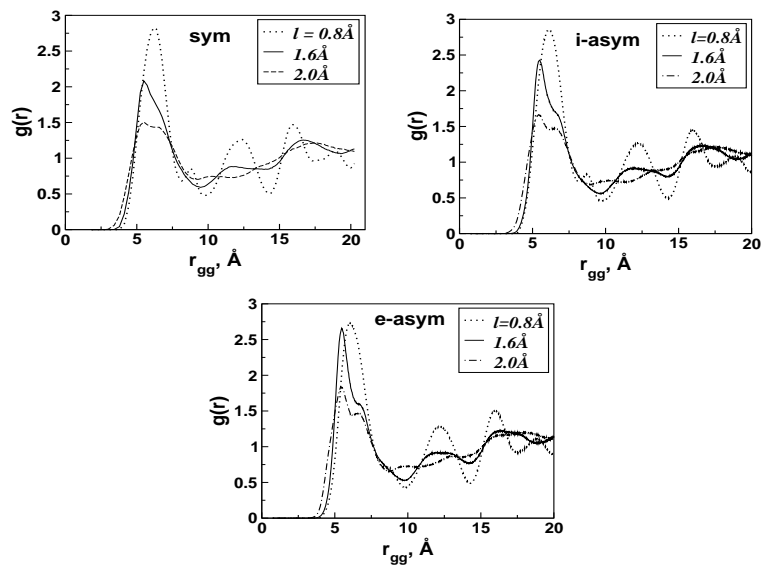}}
\caption{ Guest-guest radial distribution for a few guest sizes, $l$.
The radial distribution function reported is between the molecular center of masses
for sym, i-asym and e-asym (see text).}
\label{rdf_gg_sym}
\end{center}
\end{figure}

\begin{figure}
\begin{center}
{\includegraphics*[width=10cm]{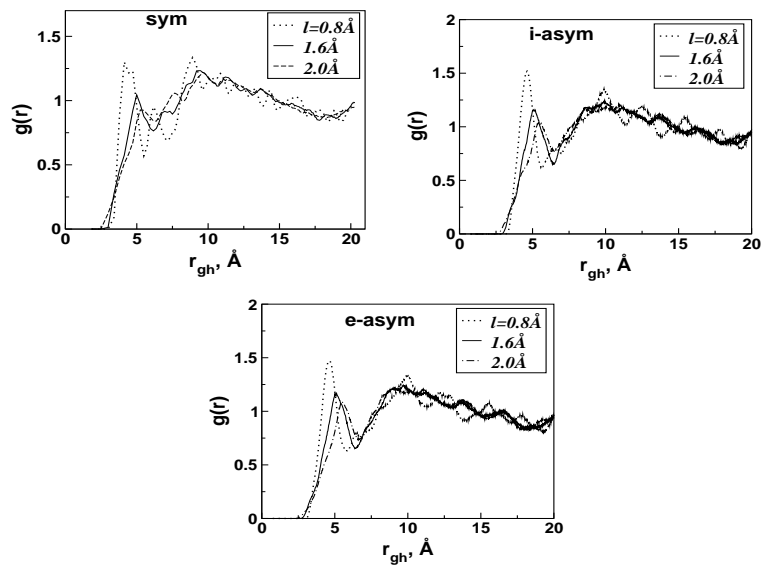}}
\caption{ Center of mass of the guest-host radial distribution function, $g_{com-h}(r)$ 
for a few guest sizes, $l$, for sym, i-asym and e-asym.}
\label{rdf_gh_sym}
\end{center}
\end{figure}

\begin{figure}
\begin{center}
{\includegraphics*[width=10cm]{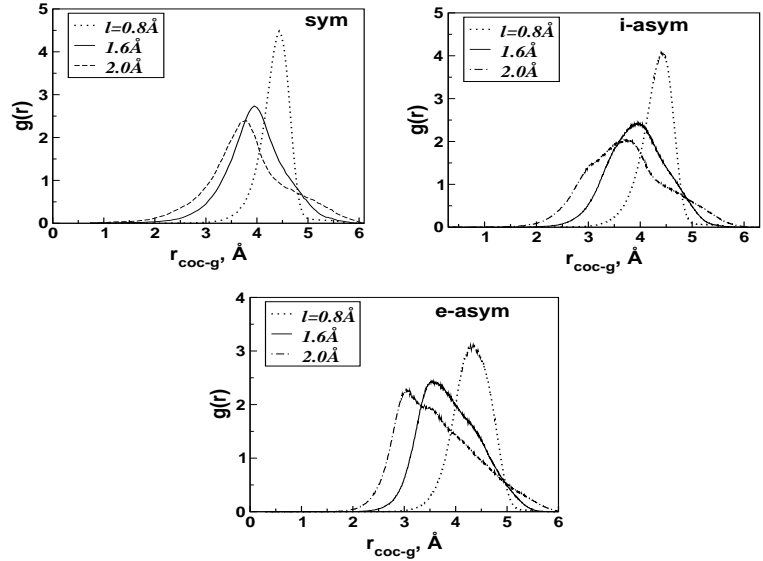}}
\caption{ The center of cage to center of mass molecule
radial distribution function. With increase in $l$, the distribution maximum shifts towards
the cage center for all the three degrees of asymmetry in interaction (sym, i-asym and e-asym).}
\label{coc_g_rdf_sym}
\end{center}
\end{figure}

The guest-guest as well as guest-zeolite radial distribution functions(rdfs) 
for different guest sizes, $l$ are shown in Figure \ref{rdf_gg_sym} 
and \ref{rdf_gh_sym} for different
degrees of asymmetry in interaction. The guest-zeolite rdfs have been 
computed between guest center of mass
and the oxygens of the zeolites. We see that for $l$ = 0.8\AA\ both the rdfs
exhibit structure and well defined peaks suggesting a more solid-like 
behaviour. In contrast,
the larger sizes $l$ = 1.6 and 2.0\AA\ exhibit a more fluid-like rdf with 
less structure.
The center of cage-center of mass radial distribution is shown in 
Figure \ref{coc_g_rdf_sym}.
This gives the radial distribution of the molecular center of mass of the 
diatomic species within
the cage. For $l$ = 0.8\AA\ we see that the molecule is close to the periphery 
of the cage.
The molecule does not occupy the central portion of the cage at all. In contrast,
the diatomic species with larger $l$ exhibit a distribution maximum at smaller $r$ 
values suggesting
that they are closer to the cage center. In addition the distribution 
for $l$ = 0.8\AA\ 
is narrow as compared to $l$ = 1.6 and 2.0\AA\ suggesting possible 
localization. This is consistent
with the solid-like rdfs we see for guest-guest and guest-zeolite rdfs. We 
shall see that the
computed self diffusivities are consistent with this.

\begin{figure}
\begin{center}
{\includegraphics*[width=11cm]{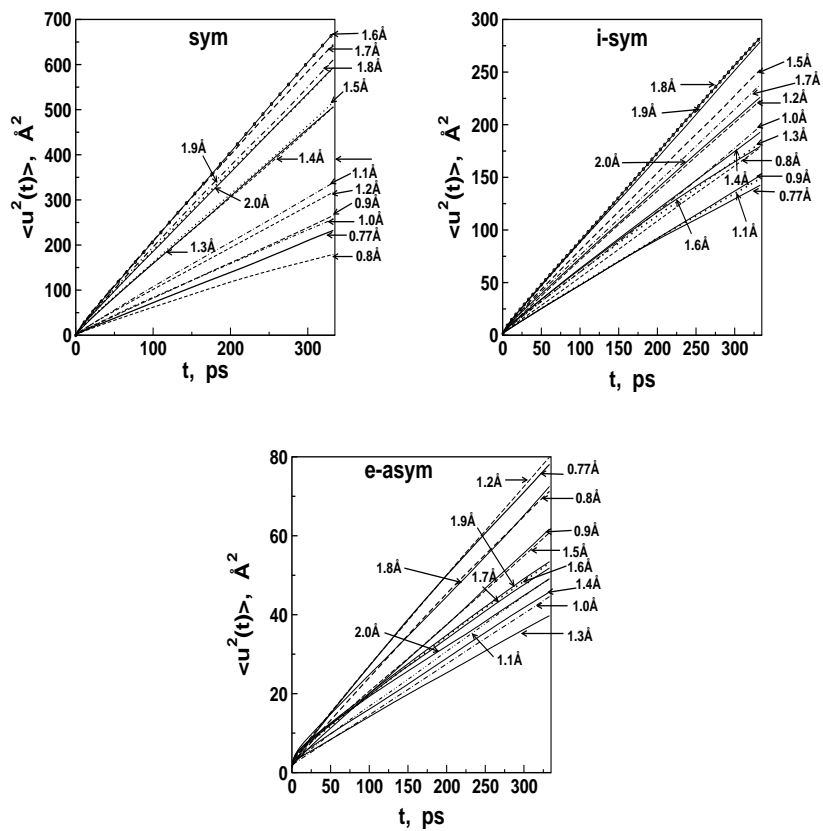}}
\caption{Time evolution of mean square displacement of guests for different values of $l$
and the three sets involving different degrees of asymmetry in interaction : sym, i-asym and
e-asym.}
\label{msd_sym}
\end{center}
\end{figure}

The mean square displacement (MSD) for guests with different bond lengths are reported
in Figure \ref{msd_sym}. The curves are straight which suggests good statistics. 
In symmetric case, there is a decrease in slope with increase in $l$ for small 
guest sizes. A gradual 
increase in the slope of MSD is observed for intermediate guest sizes with a 
maximum slope for 
$l$ = 1.6\AA. Similar, behavior is seen for i-asym case, with maximum slope for
$l$ = 1.8\AA. In case of e-asym, the slope  does not exhibit an increase 
with increase in $l$ and no 
anomalous maximum in slope is observed for any of the guest lengths. For a 
given guest size, $l$, 
the magnitude of the slope is least for guests of the e-asym or extreme 
asymmetric potential.

\begin{figure}
\begin{center}
{\includegraphics*[width=10cm]{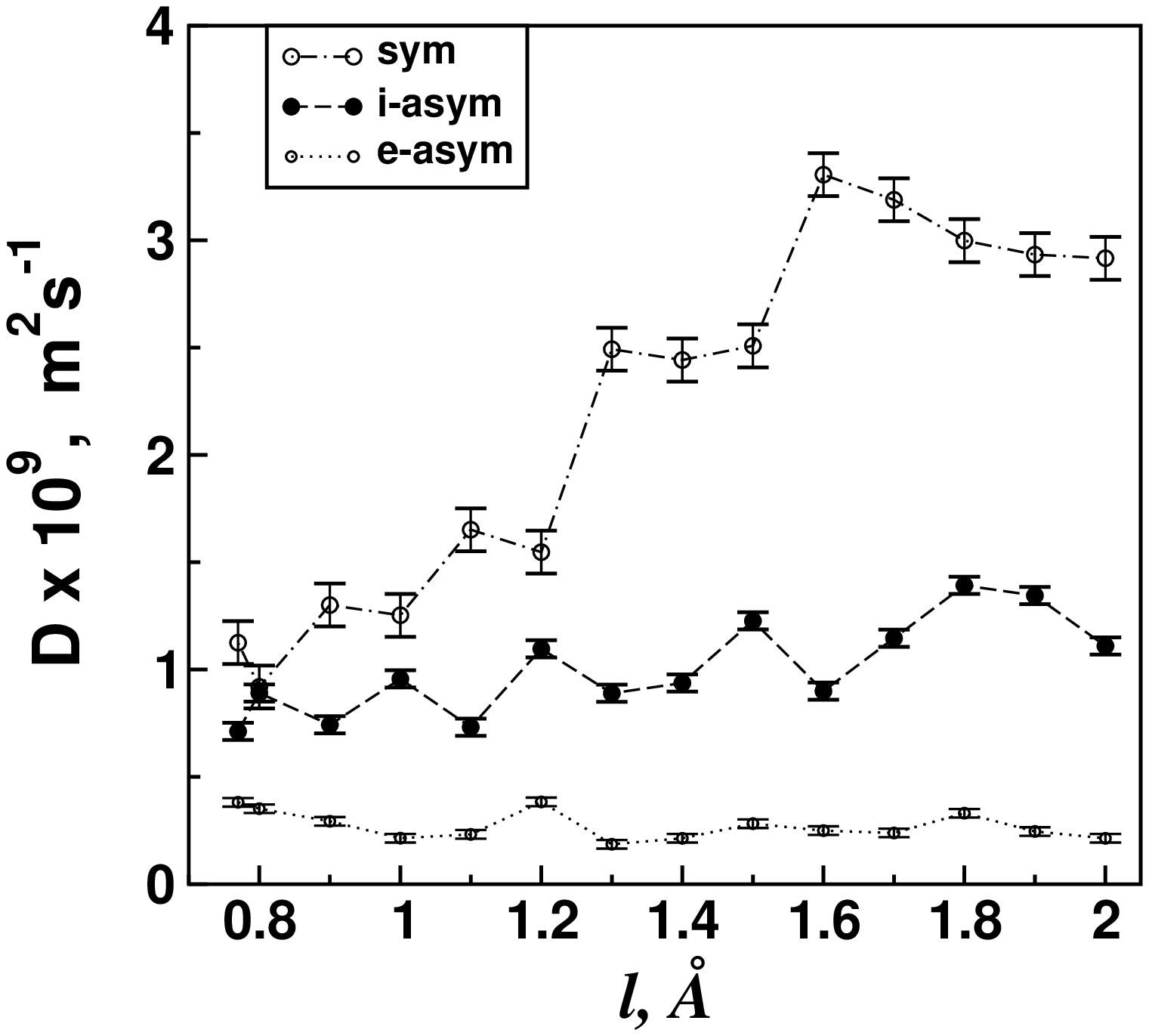}}
\caption{Diffusivity of the guest molecule as a function of the length, $l$
in zeolite-Y at 200K in presence of different types of guest-host interaction; 
sym, i-asym and e-asym cases.}
\label{D-l_sym}
\end{center}
\end{figure}

We show in Figure \ref{D-l_sym} the variation in self diffusivity as a function 
of the bond length of the diatomic molecule for the three different cases 
(a) sym (b) i-asym and (c) e-asym. Note the increase in $D$ with $l$ in 
the range 0.77-1.6\AA\ for the symmetric interaction. 
For still larger $l$, self diffusivity decreases. The
maximum at $l$ = 1.6\AA\ is seen only for the symmetric case. 
Another feature that is worth noting from the figure is the oscillating 
behavior of $D$ with $l$ : we see that $D$ increases significantly on 
going from 0.8 to 0.9\AA\ or 1.0 to 1.1\AA\ or 1.2 to 1.3\AA. Somewhat 
similar behavior has been seen in the case of window effect where the 
self diffusivity of various alkanes C$_n$H$_{2n+2}$ as a function of 
$n$ exhibit oscillatory behavior in $D$ with $n$ \cite{Gorring}.

The most important result is that as soon as asymmetry is introduced 
in the interactions between the guest and the host, the maximum in $D$ 
vanishes almost entirely. For the large asymmetry case, there is no 
increase in $D$ at all with increase in $l$. In fact, a slight decrease 
is seen upto 1.0\AA\ before leveling off. 

We note that Levitation effect has previously been noticed for monoatomic 
guests in zeolites, dense liquids and dense solids as well as for 
pentanes in zeolites. However, the present study is the first study
to investigate the role of symmetry of interaction on the diffusivity 
maximum as a consequence of the Levitation Effect. Better understanding of 
the effect and the role of symmetry will provide additional insights. 

\begin{figure}
\begin{center}
{\includegraphics*[width=10cm]{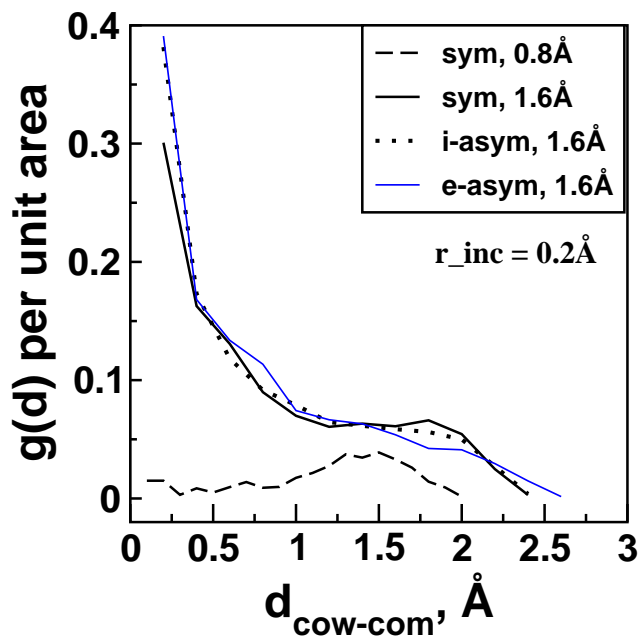}}
\caption{The distribution of center of mass of guest from the center of window
when the guest is in the plane of window for guest lengths: 0.8\AA\ and 1.6\AA(sym)
and 1.6\AA(i-asym and e-asym).}  
\label{cow-com_sym}
\end{center}
\end{figure}

Figure \ref{cow-com_sym} shows a plot of distribution, f(cow-com) of the 
distance between the center of mass from the center of window, $d_{cow-com}$ when the 
diatomic molecule is in the plane of the window. When the distance between 
the diatomic molecule and the window plane is zero, the molecule is 
 exiting from one of the $\alpha$-cages and entering a neighbouring 
$\alpha$-cage. For the symmetric case, the distribution exhibits a 
maximum near $d_{cow-com}$ = 1.5\AA\ for the diatomic from the linear 
regime while the maximum in the distribution is near zero ($d_{cow-com}$=0\AA)
for the 
diatomic at the diffusivity maximum. The coincidence of the center of 
mass of the diatomic from anomalous regime with the window 
center ($d_{cow-com}$=0\AA) leads 
to diffusivity maximum since at the window center alone there is an
inversion symmetry. The presence of inversion symmetry is required for 
mutual cancellation of force that is responsible for the diffusivity 
maximum. Thus, the presence of diffusivity maximum is associated
with the presence of an inversion symmetry and the passage of the diffusant 
through the point at which inversion symmetry exists. Here we must emphasize 
that the coincidence of the center of mass of the diatomic species with the 
window center is meant only in the statistical sense. 

For i-asym as well as e-asym cases, we see that the maximum is i
close to $d_{cow-com}$ = 0\AA. 
Now, although similar 
coincidence with the center of window (which is also a crystallographic 
center of inversion)
occurs, the asymmetry of interaction with the zeolite ensures that such a 
cancellation of forces does not occur. As a result no maximum in self 
diffusivity is seen when the bond length $d$ is similar to the window diameter.
The asymmetry arises because $\epsilon_{Ah}$ $\neq$ $\epsilon_{Bh}$ for $h$ = O, Na. 

\begin{figure}
\begin{center}
{\includegraphics*[width=10cm]{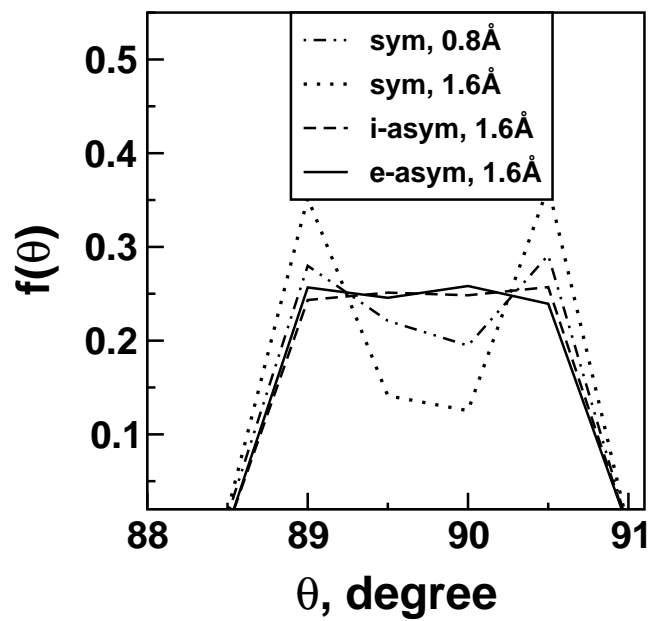}}
\caption{The distribution of angle between the unit vector 
perpendicular to the window plane and the molecular
axis of the guest molecule: 0.8 and 1.6\AA(sym), 1.6\AA(i-asym and e-asym).}  
\label{angdistr_sym}
\end{center}
\end{figure}

The angle between the molecular axis of the diatomic molecule and the 
unit vector perpendicular to the window plane when the molecular center 
of mass coincides with the window plane is shown in Figure \ref{angdistr_sym}. 
We see that the angle is nearly 90$^{\circ}$ and plane of the molecule
coincides with the plane of the window suggesting that the molecule 
prefers to go with its long axis parallel to the window plane rather 
than perpendicular. This is because this optimizes the 
interaction between the diatomic molecule and the zeolite better than when
the molecular axis is perpendicular to the window plane.

\begin{figure}
\begin{center}
{\includegraphics*[width=11cm]{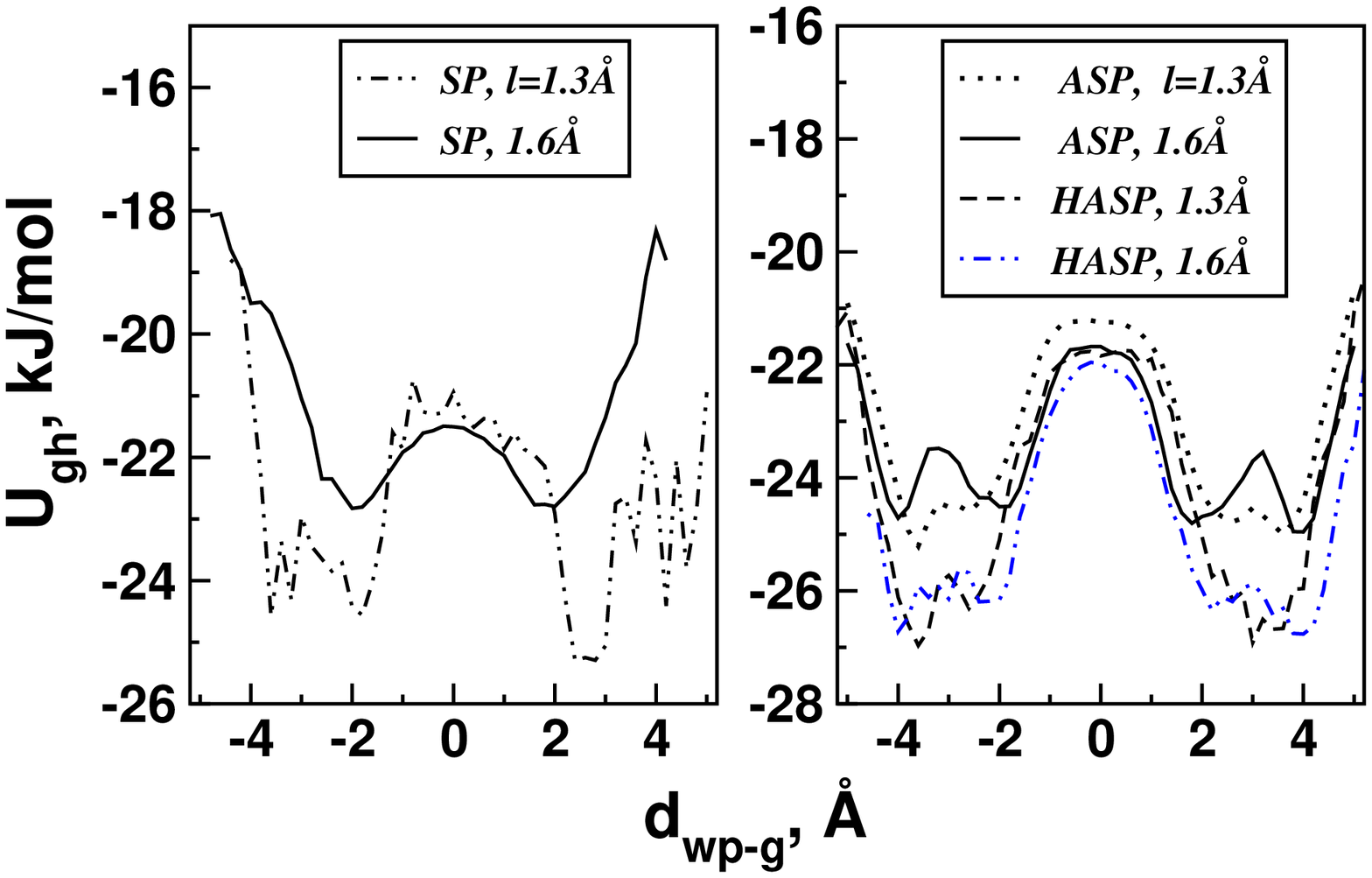}}
\caption{Variation of interaction energy of the guest with the host
as a function of the distance of the guest from the window plane, d$_wp$. The
energy profiles are shown for 1.3, 1.6 and 1.7\AA\ guest lengths for symmetric
case, 1.3 and 1.6\AA\ for i-asym and e-asym cases.}
\label{eprofile_sym}
\end{center}
\end{figure}

The activation energy barrier at the window center is a characteristic 
feature of the diffusant from the linear regime while the diatomic 
from the anomalous regime encounters either a lower barrier or even a 
negative barrier. Figure \ref{eprofile_sym} shows the variation of the 
guest-zeolite interaction energy as a function of the distance from the 
window plane. We see that for the symmetric interaction there is 
significant barrier at $d_{wg}$ = 0 for 1.3\AA\ sized guest. But for 
the larger size of 1.6 or 1.7\AA\ the barrier is lower. For i-asym as 
well as e-asym simulations, we see that the barrier is larger for 1.3 as 
well as 1.6\AA.

\begin{figure}
\begin{center}
{\includegraphics*[width=10cm]{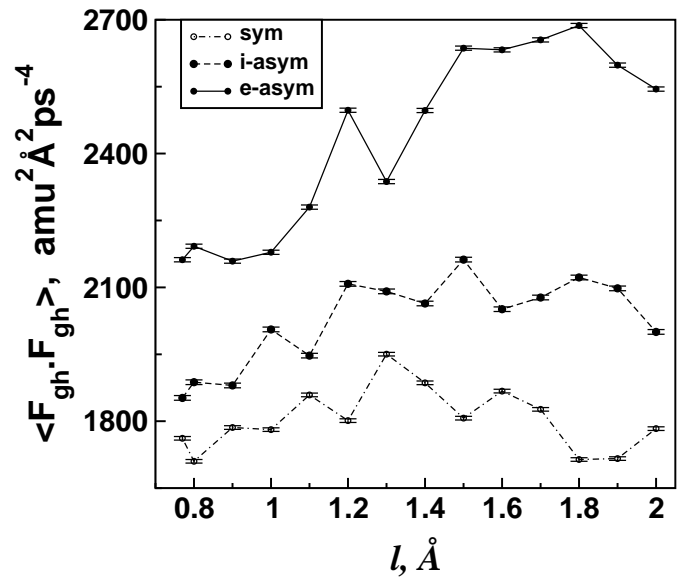}}
\caption{ Average mean square force as a function of guest length, $l$
for sym, i-asym and e-asym cases. The average mean square force is the
center of mass force of the guest due to the host atoms.}  
\label{force_sym}
\end{center}
\end{figure}

The average mean square force exerted on the guest by the zeolite averaged 
over all the molecular dynamics trajectory and all guests is shown 
in Figure \ref{force_sym} 
plotted as a function of $l$. We see that the average 
mean square force is a minimum for the $l$ for which self diffusivity 
$D$ is maximum in the case of symmetric interaction. However, with the 
introduction of asymmetry in the interaction, we note that the average 
mean square force is no more a minimum for $l$ = 1.6\AA. The diffusivity 
maximum also disappears for i-asym and e-asym sets.

\begin{figure}
\begin{center}
{\includegraphics*[width=10cm]{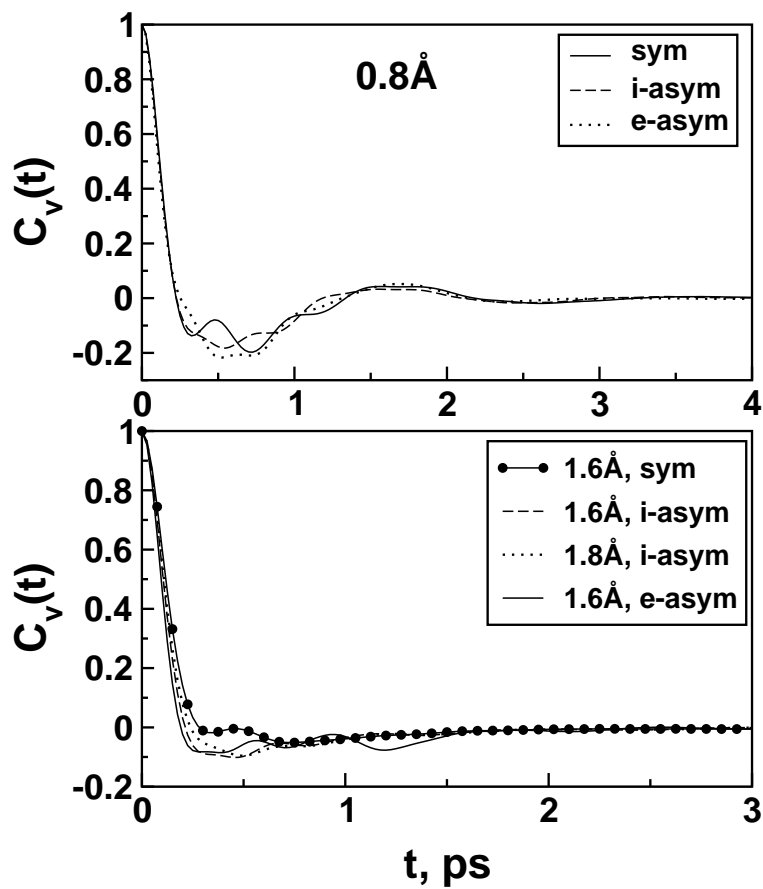}}
\caption{ Velocity autocorrelation function of guests of different $l$ diffusing
in zeolite Y at 200K in presence of different guest-host interaction, sym, i-asym
and e-asym cases.}  
\label{vcf_sym}
\end{center}
\end{figure}

The velocity autocorrelation function (VACF) for few of the guest sizes 
for the three different cases; sym, i-asym and e-asym are shown in 
Figure \ref{vcf_sym}.
The linear regime guest, 0.8\AA\ shows oscillatory behavior irrespective 
of the nature of
the interaction potential. The VACF of guest with diffusivity maximum in sym-case (1.6\AA) 
has less backscattering as compared to the 1.6\AA\ guest length in case of 
i-asym and e-asym cases and linear regime guest in all the cases. 
The guest with maximum diffusivity in case of sym-potential has a lower  
energy barrier at the window leading to its facile passage past the window. 
Linear regime guest has a higher energy barrier at the window
therefore has negative correlation in the VACF. 
With increase in asymmetry in the guest-host potential, diffusivity maximum 
for intermediate
guest sizes disappears which is seen with an oscillatory behavior of VACF 
for the intermediate
guest sizes in case of i-asym and e-asym potential.
 
\begin{figure}
\begin{center}
{\includegraphics*[width=10cm]{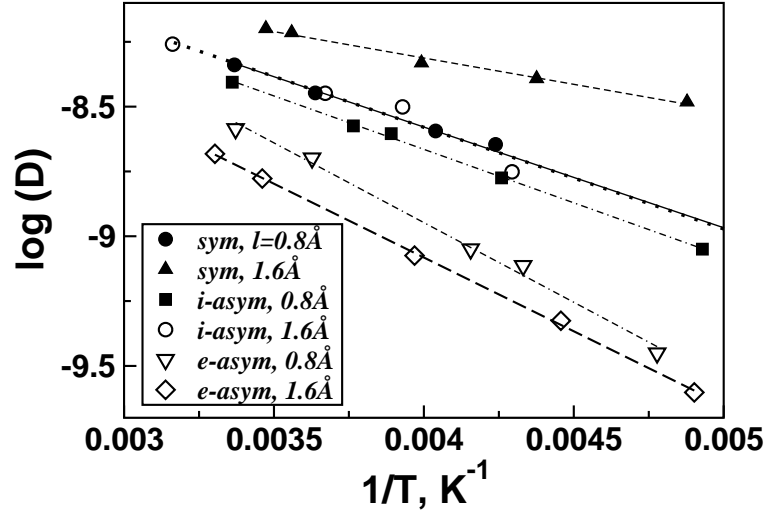}}
\caption{Arrhenius plot of the different guest lengths, $l$ for 
case, sym, i-asym
and e-asym obtained from the diffusivities at average temperatures (obtained
from the simulation runs performed at desired temperatures of 200, 225, 
250, 275 and 300K).}
\label{arrhen_sym}
\end{center}
\end{figure}

Figure \ref{arrhen_sym} shows the Arrhenius plots  for
few of the guest sizes; sym case (0.8, 1.6\AA), i-asym 
case (0.8, 1.6\AA) and 
e-asym case (0.8, 1.6\AA) between $ln(D)$ and inverse of average
temperatures obtained from the simulation runs. 
The diffusivities for the Arrhenius plot have been obtained by simulating
the systems at five different temperatures (200, 225, 250, 275 
and 300K). The activation energies 
for diffusion obtained 
from the slope of Arrhenius plot are shown in Table \ref{Eact_sym}. The activation energy is 
minimum for the guest length 1.6\AA\ for symmetric interaction. 
This is also the guest which shows maximum diffusivity. Such a trend is 
not observed in i-asym and e-asym cases.

\begin{table}
\caption {Activation energies of guests of different $l$ obtained from Arrhenius plot
for different cases: sym, i-asym and e-asym.}
\begin{center}
\begin{tabular}{|c|c|c|}\hline
{case}&{$l$, \AA} & {E$_{a}$, kJ/mol} \\\hline
sym&0.8&7.4358 \\
sym&1.6&3.8927  \\
i-asym&0.8&7.8917 \\
i-asym&1.6&7.5054 \\
e-asym&0.8&11.7714 \\
e-asym&1.6&10.8939 \\ \hline
\end{tabular}
\label{Eact_sym}
\end{center}
\end{table}

\section{Conclusions}

In summary, the present work reports results for diatomic species AB in zeolite NaY. The results
show that when A = B, there is a maximum in the self diffusivity of the diatomic molecule
at large bond lengths $d_{AB}$. This suggests that Levitation Effect exists for diatomic species
AA. Results when A $\neq$ B with small asymmetry in the interaction between A and B with the
atoms of the zeolite shows a weak maximum in the self diffusivity. When the interaction strength
between A and host atoms are made very different from that of B with host atoms, it is seen that
such a maximum in $D$ disappears completely. This suggests the absence of Levitation Effect for
such a system.

A few remarks on the symmetry necessary for the diffusivity maximum or Levitation Effect is worth noting.
The inversion symmetry which is essential is not the crystallographically defined symmetry which is
based on structure. The necessary symmetry we require for ensuring that the Levitation Effect is seen
is interaction inversion symmetry. Interaction inversion symmetry requires that the force on the diffusant
from given direction is equal and opposite to the force from the diagonally opposite direction.
This is less stringent a requirement than the crystallographic inversion symmetry. 
The latter, however,
ensures the existence of interaction inversion symmetry. Situations where there 
is no crystallographic
inversion symmetry but there is interaction inversion symmetry are when the 
forces arising from
atoms at different distances add upto along a given direction to say, {\bf F$_p$}.
Now although the atomic arrangement along the diagonally opposite direction is completely different,
equality should be seen along all directions.

\begin{figure}
\begin{center}
{\includegraphics*[width=10cm]{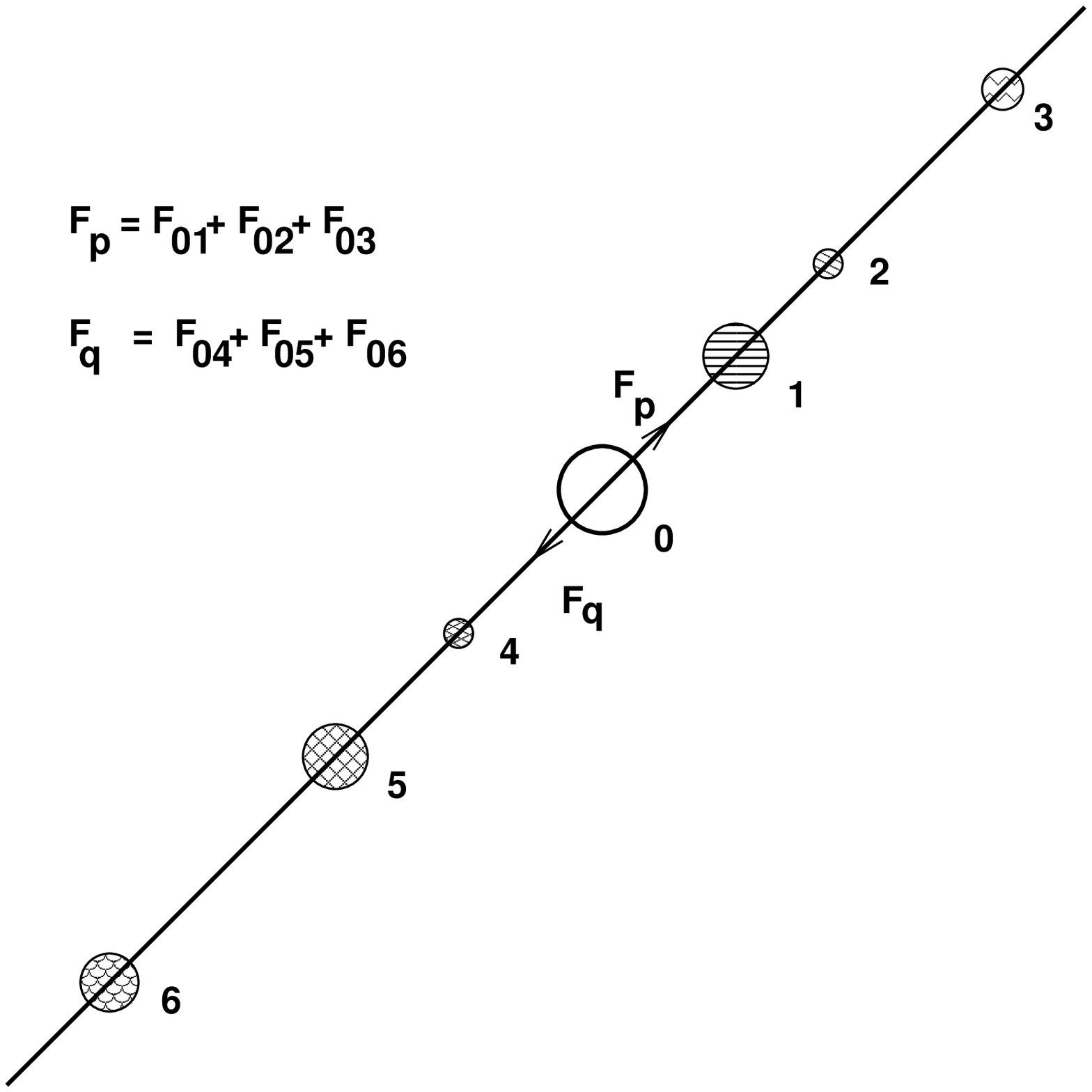}}
\caption{Schematic diagram of equal and opposite forces due to 
interaction inversion symmetry on center atoms occurring due to the 
neighboring atoms.}
\label{iis1_sym}
\end{center}
\end{figure}

These results suggest the role of symmetry in the interaction, leading to Levitation Effect.
The present work demonstrates the need for symmetry in the interaction to observe the diffusivity maximum
and Levitation Effect. Absence of symmetry in the interaction leads to obliteration of the
diffusivity maximum or Levitation Effect.

\noindent
{\em Acknowledgment} : Authors wish to thank Department of Science and
Technology, New Delhi and CSIR, New Delhi, for financial support in carrying out this work.
Authors also acknowledge C.S.I.R., New Delhi for a research fellowship to
M.S.

\bibliography{manju_symm}
\bibliographystyle{plain}

\end{document}